\documentclass{icrc29}
\usepackage{graphicx,amssymb,amsmath,times}
\usepackage{latexsym}

\newcommand{\cc}{cosmological constant}
\newcommand{\del}{\partial}
\newcommand{\dphi}{\partial_i \phi \partial^i \phi}

\begin{document}

\title[Darker Side of the Universe]{ Darker Side of the Universe\\
....{\it and the crying need for some bright ideas!}}

\author[T. Padmanabhan]{T.~Padmanabhan\\
IUCAA, 
Post Bag 4, Ganeshkhind,\\
 Pune - 411 007, INDIA.\\
email: nabhan@iucaa.ernet.in}

\presenter{Presenter: T. Padmanabhan}

\maketitle

\begin{abstract}

Observations suggest that nearly seventy  per cent of the energy density in the universe is unclustered and exerts negative pressure. 
Theoretical understanding of this component (`dark energy'), which is driving an accelerated expansion of the universe, is {\it the}
problem in cosmology today. I discuss this issue with  special emphasis on the cosmological constant as the possible choice for the dark energy. Several curious  features of a universe with a cosmological constant are
described and some possible approaches to understand the nature of the cosmological constant are discussed. 

 \end{abstract}


\section{The Contents of the Cosmos}

 Data of exquisite quality which became available in the last couple of decades has confirmed the broad paradigm of standard
 cosmology and has helped us to determine the composition of  the universe. As a direct consequence, these cosmological observations have thrusted upon us a rather preposterous
composition for the universe which defies any simple explanation, thereby posing the greatest challenge
theoretical physics has ever faced.

It is convenient to measure
the energy densities of the various components in terms of a \textit{critical energy density} $\rho_c=3H^2_0/8\pi G$  where $H_0=(\dot a/a)_0$
is the rate of expansion of the universe at present.  The variables $\Omega_i=\rho_i/\rho_c$ 
will give the fractional contribution of different components of the universe ($i$ denoting baryons, dark matter, radiation, etc.) to the  critical density. Observations then lead to the following results:

\begin{itemize}
\item
Our universe has $0.98\lesssim\Omega_{tot}\lesssim1.08$. The value of $\Omega_{tot}$ can be determined from the angular anisotropy spectrum of the cosmic microwave background radiation (CMBR) (with the reasonable assumption that $h>0.5$) and these observations now show that we live in a universe
with critical density \cite{cmbr}. 
\item
Observations of primordial deuterium produced in big bang nucleosynthesis (which took place when the universe
was about 1 minute in age) as well as the CMBR observations show that  \cite{baryon} the {\it total} amount of baryons in the
universe contributes about $\Omega_B=(0.024\pm 0.0012)h^{-2}$. Given the independent observations on the Hubble constant \cite{h} which fix $h=0.72\pm 0.07$, we conclude that   $\Omega_B\cong 0.04-0.06$. These observations take into account all baryons which exist in the universe today irrespective of whether they are luminous or not. Combined with previous item we conclude that
most of the universe is non-baryonic.
\item
Host of observations related to large scale structure and dynamics (rotation curves of galaxies, estimate of cluster masses, gravitational lensing, galaxy surveys ..) all suggest \cite {dm} that the universe is populated by a non-luminous component of matter (dark matter; DM hereafter) made of weakly interacting massive particles which \textit{does} cluster at galactic scales. This component contributes about $\Omega_{DM}\cong 0.20-0.35$.
\item
Combining the last observation with the first we conclude that there must be (at least) one more component 
to the energy density of the universe contributing about 70\% of critical density. Early analysis of several observations
\cite{earlyde} indicated that this component is unclustered and has negative pressure. This is confirmed dramatically by the supernova observations (see \cite{sn}; for a critical look at the data, see \cite{tptirthsn1,jbp}).  The observations suggest that the missing component has 
$w=p/\rho\lesssim-0.78$
and contributes $\Omega_{DE}\cong 0.60-0.75$.
\item
The universe also contains radiation contributing an energy density $\Omega_Rh^2=2.56\times 10^{-5}$ today most of which is due to
photons in the CMBR. This is dynamically irrelevant today but would have been the dominant component in the universe  at redshifts
larger that $z_{eq}\simeq \Omega_{DM}/\Omega_R\simeq 4\times 10^4\Omega_{DM}h^2$.
\item
Together we conclude that our universe has (approximately) $\Omega_{DE}\simeq 0.7,\Omega_{DM}\simeq 0.26,\Omega_B\simeq 0.04,\Omega_R\simeq 5\times 10^{-5}$. All known observations
are consistent with such an --- admittedly weird --- composition for the universe.
\end{itemize}

Before discussing the puzzles raised by the composition of the universe in greater detail, let us briefly remind ourselves of the \textit{successes} of the standard paradigm. The  key idea is that if there existed small fluctuations in the energy density in the early universe, then gravitational instability can amplify them in a well-understood manner  leading to structures like galaxies etc. today. The most popular model for generating these fluctuations is based on the idea that if the very early universe went through an inflationary phase \cite{inflation}, then the quantum fluctuations of the field driving the inflation can lead to energy density fluctuations\cite{genofpert,tplp}. It is possible to construct models of inflation such that these fluctuations are described by a Gaussian random field and are characterized by a power spectrum of the form $P(k)=A k^n$ with $n\simeq 1$. The models cannot predict the value of the amplitude $A$ in an unambiguous manner but it can be determined from CMBR observations. The CMBR observations are consistent with the inflationary model for the generation of perturbations and gives $A\simeq (28.3 h^{-1} Mpc)^4$ and $n=0.97\pm0.023$ (The first results were from COBE \cite{cobeanaly} and
WMAP has reconfirmed them with far greater accuracy).
When the perturbation is small, one can use well defined linear perturbation theory to study its growth. But when $\delta\approx(\delta\rho/\rho)$ is comparable to unity the perturbation theory
breaks down. Since there is more power at small scales, smaller scales go non-linear first and structure forms hierarchically. 
The non linear evolution of the  \textit{dark matter halos} (which is an example of statistical mechanics
 of self gravitating systems; see e.g.\cite{smofgs}) can be understood by simulations as well as theoretical models based on approximate ansatz
\cite{nlapprox} and  nonlinear scaling relations \cite{nsr}.
 The baryons in the halo will cool and undergo collapse
 in a fairly complex manner because of gas dynamical processes. 
 It seems unlikely that the baryonic collapse and galaxy formation can be understood
 by analytic approximations; one needs to do high resolution computer simulations
 to make any progress \cite{baryonsimulations}. 
 
 All these results are broadly consistent with observations. 
 So, to the zeroth order, the universe is characterized by just seven numbers: $h\approx 0.7$ describing the current rate of expansion; $\Omega_{DE}\simeq 0.7,\Omega_{DM}\simeq 0.26,\Omega_B\simeq 0.04,\Omega_R\simeq 5\times 10^{-5}$ giving the composition of the universe; the amplitude $A\simeq (28.3 h^{-1} Mpc)^4$ and the index $n\simeq 1$ of the initial perturbations. 
The challenge is to make some sense out of these numbers from a more fundamental point of view.

\section{The Dark Energy}

It is rather frustrating that the only component of the universe which we understand theoretically is the radiation! While understanding the
baryonic and dark matter components [in particular the values of $\Omega_B$ and $\Omega_{DM}$] is by no means trivial, the issue of dark energy is lot more perplexing, thereby justifying the attention it has received recently.

The key observational feature of dark energy is that --- treated as a fluid with a stress tensor $T^a_b=$ dia     $(\rho, -p, -p,-p)$ 
--- it has an equation state $p=w\rho$ with $w \lesssim -0.8$ at the present epoch. 
The spatial part  ${\bf g}$  of the geodesic acceleration (which measures the 
  relative acceleration of two geodesics in the spacetime) satisfies an \textit{exact} equation
  in general relativity  given by:
  \begin{equation}
  \nabla \cdot {\bf g} = - 4\pi G (\rho + 3p)
  \label{nextnine}
  \end{equation} 
 This  shows that the source of geodesic  acceleration is $(\rho + 3p)$ and not $\rho$.
  As long as $(\rho + 3p) > 0$, gravity remains attractive while $(\rho + 3p) <0$ can
  lead to repulsive gravitational effects. In other words, dark energy with sufficiently negative pressure will
  accelerate the expansion of the universe, once it starts dominating over the normal matter.  This is precisely what is established from the study of high redshift supernova, which can be used to determine the expansion
rate of the universe in the past \cite{sn}. 

The simplest model for  a fluid with negative pressure is the
cosmological constant (for a review, see \cite{cc}) with $w=-1,\rho =-p=$ constant.
If the dark energy is indeed a cosmological constant, then it introduces a fundamental length scale in the theory $L_\Lambda\equiv H_\Lambda^{-1}$, related to the constant dark energy density $\rho_{_{\rm DE}}$ by 
$H_\Lambda^2\equiv (8\pi G\rho_{_{\rm DE}}/3)$.
In classical general relativity,
    based on the constants $G, c $ and $L_\Lambda$,  it
  is not possible to construct any dimensionless combination from these constants. But when one introduces the Planck constant, $\hbar$, it is  possible
  to form the dimensionless combination $H^2_\Lambda(G\hbar/c^3) \equiv  (L_P^2/L_\Lambda^2)$.
  Observations then require $(L_P^2/L_\Lambda^2) \lesssim 10^{-123}$.
  As has been mentioned several times in literature, this will require enormous fine tuning. What is more,
 in the past, the energy density of 
  normal matter and radiation  would have been higher while the energy density contributed by the  cosmological constant
  does not change.  Hence we need to adjust the energy densities
  of normal matter and cosmological constant in the early epoch very carefully so that
  $\rho_\Lambda\gtrsim \rho_{\rm NR}$ around the current epoch.
  This raises the second of the two cosmological constant problems:
  Why is it that $(\rho_\Lambda/ \rho_{\rm NR}) = \mathcal{O} (1)$ at the 
  {\it current} phase of the universe ?

  Because of these conceptual problems associated with the cosmological constant, people have explored a large variety of alternative possibilities. The most popular among them uses a scalar field $\phi$ with a suitably chosen potential $V(\phi)$ so as to make the vacuum energy vary with time. The hope then is that, one can find a model in which the current value can be explained naturally without any fine tuning.
  A simple form of the source with variable $w$ are   scalar fields with
  Lagrangians of different forms, of which we will discuss two possibilities:
    \begin{equation}
  L_{\rm quin} = \frac{1}{2} \partial_a \phi \partial^a \phi - V(\phi); \quad L_{\rm tach}
  = -V(\phi) [1-\partial_a\phi\partial^a\phi]^{1/2}
  \label{lquineq}
  \end{equation}
  Both these Lagrangians involve one arbitrary function $V(\phi)$. The first one,
  $L_{\rm quin}$,  which is a natural generalization of the Lagrangian for
  a non-relativistic particle, $L=(1/2)\dot q^2 -V(q)$, is usually called quintessence (for
  a small sample of models, see \cite{phiindustry}; there is an extensive and growing literature on scalar field models and more references can be found in the reviews in ref.\cite{cc}).
    When it acts as a source in Friedman universe,
   it is characterized by a time dependent
  $w(t)$ with
    \begin{equation}
  \rho_q(t) = \frac{1}{2} \dot\phi^2 + V; \quad p_q(t) = \frac{1}{2} \dot\phi^2 - V; \quad w_q
  = \frac{1-(2V/\dot\phi^2)}{1+ (2V/\dot\phi^2)}
  \label{quintdetail}
  \end{equation}

The structure of the second Lagrangian  in Eq.~(\ref{lquineq}) can be understood by a simple analogy from
special relativity. A relativistic particle with  (one dimensional) position
$q(t)$ and mass $m$ is described by the Lagrangian $L = -m \sqrt{1-\dot q^2}$.
It has the energy $E = m/  \sqrt{1-\dot q^2}$ and momentum $k = m \dot
q/\sqrt{1-\dot q^2} $ which are related by $E^2 = k^2 + m^2$.  As is well
known, this allows the possibility of having \textit{massless} particles with finite
energy for which $E^2=k^2$. This is achieved by taking the limit of $m \to 0$
and $\dot q \to 1$, while keeping the ratio in $E = m/  \sqrt{1-\dot q^2}$
finite.  The momentum acquires a life of its own,  unconnected with the
velocity  $\dot q$, and the energy is expressed in terms of the  momentum
(rather than in terms of $\dot q$)  in the Hamiltonian formulation. We can now
construct a field theory by upgrading $q(t)$ to a field $\phi$. Relativistic
invariance now  requires $\phi $ to depend on both space and time [$\phi =
\phi(t, {\bf x})$] and $\dot q^2$ to be replaced by $\partial_i \phi \partial^i
\phi$. It is also possible now to treat the mass parameter $m$ as a function of
$\phi$, say, $V(\phi)$ thereby obtaining a field theoretic Lagrangian $L =-
V(\phi) \sqrt{1 - \del^i \phi \del_i \phi}$. The Hamiltonian  structure of this
theory is algebraically very similar to the special  relativistic example  we
started with. In particular, the theory allows solutions in which $V\to 0$,
$\dphi \to 1$ simultaneously, keeping the energy (density) finite.  Such
solutions will have finite momentum density (analogous to a massless particle
with finite  momentum $k$) and energy density. Since the solutions can now
depend on both space and time (unlike the special relativistic example in which
$q$ depended only on time), the momentum density can be an arbitrary function
of the spatial coordinate. The structure of this Lagrangian is similar to those analyzed in a wide class of models
   called {\it K-essence} \cite{kessence} and  provides a rich gamut of possibilities in the
context of cosmology
 \cite{tptachyon,tachyon}.

   Since  the quintessence field (or the tachyonic field)   has
   an undetermined free function $V(\phi)$, it is possible to choose this function
  in order to produce a given $H(a)$.
  To see this explicitly, let
   us assume that the universe has two forms of energy density with $\rho(a) =\rho_{\rm known}
  (a) + \rho_\phi(a)$ where $\rho_{\rm known}(a)$ arises from any known forms of source 
  (matter, radiation, ...) and
  $\rho_\phi(a) $ is due to a scalar field.  
  Let us first consider quintessence. Here,  the potential is given implicitly by the form
  \cite{ellis,tptachyon}
  \begin{equation}
  V(a) = \frac{1}{16\pi G} H (1-Q)\left[6H + 2aH' - \frac{aH Q'}{1-Q}\right]
  \label{voft}
   \end{equation} 
    \begin{equation}
    \phi (a) =  \left[ \frac{1}{8\pi G}\right]^{1/2} \int \frac{da}{a}
     \left[ aQ' - (1-Q)\frac{d \ln H^2}{d\ln a}\right]^{1/2}
    \label{phioft}
    \end{equation} 
   where $Q (a) \equiv [8\pi G \rho_{\rm known}(a) / 3H^2(a)]$ and prime denotes differentiation with respect to $a$.
   Given any
   $H(a),Q(a)$, these equations determine $V(a)$ and $\phi(a)$ and thus the potential $V(\phi)$. 
   \textit{Every quintessence model studied in the literature can be obtained from these equations.}
  
  Similar results exists for the tachyonic scalar field as well \cite{tptachyon}. For example, given
  any $H(a)$, one can construct a tachyonic potential $V(\phi)$ so that the scalar field is the 
  source for the cosmology. The equations determining $V(\phi)$  are now given by:
  \begin{equation}
  \phi(a) = \int \frac{da}{aH} \left(\frac{aQ'}{3(1-Q)}
   -{2\over 3}{a H'\over H}\right)^{1/2}
  \label{finalone}
  \end{equation}
   \begin{equation}
   V = {3H^2 \over 8\pi G}(1-Q) \left( 1 + {2\over 3}{a H'\over H}-\frac{aQ'}{3(1-Q)}\right)^{1/2}
   \label{finaltwo}
   \end{equation}
   Equations (\ref{finalone}) and (\ref{finaltwo}) completely solve the problem. Given any
   $H(a)$, these equations determine $V(a)$ and $\phi(a)$ and thus the potential $V(\phi)$. 
A wide variety of phenomenological models with time dependent
  \cc\ have been considered in the literature all of which can be 
   mapped to a 
  scalar field model with a suitable $V(\phi)$.

  While the scalar field models enjoy considerable popularity (one reason being they are easy to construct!)
  it is very doubtful whether they have helped us to understand the nature of the dark energy
  at any deeper level. These
  models, viewed objectively, suffer from several shortcomings:
  \begin{figure}[ht]
 \begin{center}
 \includegraphics[scale=0.5]{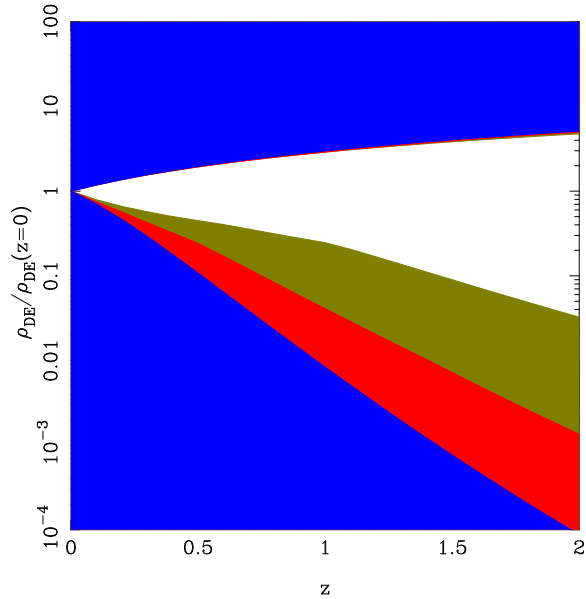}
 \end{center}
 \caption{Constraints on the possible variation of the dark energy density with redshift. The darker shaded region (blue) is excluded by SN observations while the lighter shaded region (green and red) is excluded by WMAP observations. It is obvious that WMAP puts stronger constraints on the possible
 variations of dark energy density. The cosmological constant corresponds to the horizontal line
 at unity which is consistent with observations.  (For more details, see
 the last two references in \cite{jbp}.) }
 \label{fig:bjp2ps}
 \end{figure}  
  \begin{itemize}
  \item
  They completely lack predictive power. As explicitly demonstrated above, virtually every form of $a(t)$ can be modeled by a suitable ``designer" $V(\phi)$.
  \item
  These models are  degenerate in another sense. The previous discussion  illustrates that even when $w(a)$ is known/specified, it is not possible to proceed further and determine
  the nature of the scalar field Lagrangian. The explicit examples given above show that there
  are {\em at least} two different forms of scalar field Lagrangians (corresponding to
  the quintessence or the tachyonic field) which could lead to
  the same $w(a)$. (See ref.\cite{tptirthsn1} for an explicit example of such a construction.)
  \item
  All the scalar field potentials require fine tuning of the parameters in order to be viable. This is obvious in the quintessence models in which adding a constant to the potential is the same as invoking a \cc. So to make the quintessence models work, \textit{we first need to assume the \cc\ is zero.} These models, therefore, merely push the cosmological constant problem to another level, making it somebody else's problem!.
  \item
  By and large, the potentials  used in the literature have no natural field theoretical justification. All of them are non-renormalisable in the conventional sense and have to be interpreted as a low energy effective potential in an ad hoc manner.
  \item
  One key difference between \cc\ and scalar field models is that the latter lead to a $w(a)$ which varies with time. If observations have demanded this, or even if observations have ruled out $w=-1$ at the present epoch,
  then one would have been forced to take alternative models seriously. However, all available observations are consistent with \cc\ ($w=-1$) and --- in fact --- the possible variation of $w$ is strongly constrained \cite{jbp} as shown in Figure \ref{fig:bjp2ps}. 
  \item
  While on the topic of observational constraints on $w(t)$, it must be stressed that: (a) There is fair amount of tension between WMAP and SN data  and one should be very careful about the priors used in these analysis. (b) There is no observational evidence for $w<-1$. (c) It is likely that more homogeneous, future, data sets of SN might show better agreement with WMAP results. (For more details related to these issues, see the last reference in \cite{jbp}.)
 \end{itemize}

 Given this situation, we shall now take a more serious look at the \cc\ as the source of dark energy in the universe.
 
 \section{Cosmological Constant: Facing up to the Challenge }
 
The observational and theoretical features described above suggests that one should consider \cc\ as the most natural candidate for dark energy. Though it leads to well know fine tuning problems, it also has certain attractive features that need to kept in mind.
\begin{itemize}
\item
Cosmological constant is the most economical [just one number] and simplest  explanation for all the observations. I repeat that there is absolutely \textit{no} evidence for variation of dark energy density with redshift, which is consistent with the assumption of \cc\ .
\item
Once we invoke the \cc\ classical gravity will be described by the three constants $G,c$ and $\Lambda\equiv L_\Lambda^{-2}$. It is not possible to obtain a dimensionless quantity from these; so, within classical theory, there is no fine tuning issue. Since $\Lambda(G\hbar/c^3)\equiv (L_P/L_\Lambda)^2\approx 10^{-123}$, it is obvious that the \cc\ is telling us something regarding \textit{quantum gravity}, indicated by the combination $G\hbar$. \textit{An acid test for any quantum gravity model will be its ability to explain this value;} needless to say, all the currently available models --- strings, loops etc.  --- flunk this test.
\item
So, if dark energy is indeed \cc\, this will be the greatest contribution from cosmology to fundamental physics. It will be unfortunate if we miss this chance by invoking some scalar field epicycles!
\end{itemize}

In this context, it is worth  stressing another   peculiar feature of \cc\, when it is  treated as a clue to quantum gravity.
It is well known that, based on energy scales, the \cc\ problem is an infra red problem \textit{par excellence}.
At the same time, it is a relic of a quantum gravitational effect or principle of unknown nature. An analogy will be helpful to illustrate this point. Suppose you solve the Schrodinger equation for the Helium atom for the quantum states of the two electrons $\psi(x_1,x_2)$. When the result is compared with observations, you will find that only half the states --- those in which  $\psi(x_1,x_2)$ is antisymmetric under $x_1\longleftrightarrow x_2$ interchange --- are realised in nature. But the low energy Hamiltonian for electrons in the Helium atom has no information about
this effect! Here is low energy (IR) effect which is a relic of relativistic quantum field theory (spin-statistics theorem) that is  totally non perturbative, in the sense that writing corrections to the Helium atom Hamiltonian in some $(1/c)$ expansion will {\it not} reproduce this result. I suspect the current value of \cc\ is related to quantum gravity in a similar way. There must exist a deep principle in quantum gravity which leaves its non perturbative trace even in the low energy limit
that appears as the \cc\ .

Let us now turn our attention to few of the many attempts to understand the \cc. The choice  is, of course, dictated by personal bias and is definitely a non-representative sample. A host of other approaches exist in literature, some of which can be found in \cite{catchall}.

\subsection{Gravitational Holography}

One possible way of addressing this issue is to simply eliminate from the gravitational theory those modes which couple to cosmological constant. If, for example, we have a theory in which the source of gravity is
$(\rho +p)$ rather than $(\rho +3p)$ in Eq. (\ref{nextnine}), then \cc\ will not couple to gravity at all. (The non linear coupling of matter with gravity has several subtleties; see eg. \cite{gravitonmyth}.) Unfortunately
it is not possible to develop a covariant theory of gravity using $(\rho +p)$ as the source. But we can probably gain some insight from the following considerations. Any metric $g_{ab}$ can be expressed in the form $g_{ab}=f^2(x)q_{ab}$ such that
${\rm det}\, q=1$ so that ${\rm det}\, g=f^4$. From the action functional for gravity
\begin{equation}
A=\frac{1}{16\pi G}\int d^4x (R -2\Lambda)\sqrt{-g}
=\frac{1}{16\pi G}\int d^4x R \sqrt{-g}-\frac{\Lambda}{8\pi G}\int d^4x f^4(x)
\end{equation}
it is obvious that the \cc\ couples {\it only} to the conformal factor $f$. So if we consider a theory of gravity in which $f^4=\sqrt{-g}$ is kept constant and only $q_{ab}$ is varied, then such a model will be oblivious of
direct coupling to \cc. If the action (without the $\Lambda$ term) is varied, keeping ${\rm det}\, g=-1$, say, then one is lead to a {\it unimodular theory of gravity} that has  the equations of motion 
$R_{ab}-(1/4)g_{ab}R=\kappa(T_{ab}-(1/4)g_{ab}T)$ with zero trace on both sides. Using the Bianchi identity, it is now easy to show that this is equivalent to the usual  theory with an {\it  arbitrary} \cc. That is, \cc\ arises as an undetermined integration constant in this model \cite{unimod}. 

The same result arises in another, completely different approach to gravity. In the standard approach to gravity one uses the Einstein-Hilbert Lagrangian  $L_{EH}\propto R$ which has a formal structure $L_{EH}\sim R\sim (\partial g)^2+\partial^2g$. 
If the surface term obtained by integrating $L_{sur}\propto \partial^2g$ is ignored (or, more formally, canceled by an extrinsic curvature term) then the Einstein's equations arise from the variation of the bulk
term $L_{bulk}\propto (\partial g)^2$ which is the non-covariant $\Gamma^2$ Lagrangian.
There is, however, a remarkable relation \cite{comment}
between $L_{bulk}$ and $L_{sur}$:
\begin{equation}
\sqrt{-g}L_{sur}=-\partial_a\left(g_{ij}
\frac{\partial \sqrt{-g}L_{bulk}}{\partial(\partial_ag_{ij})}\right)
\end{equation}
which allows a dual description of gravity using either $L_{bulk}$ or $L_{sur}$!
It is possible to obtain \cite{tpholo} the dynamics of gravity from an approach which  uses \textit{only} the surface term of the Hilbert action; \textit{ we do not need the bulk term at all !}. This suggests that \textit{the true  degrees of freedom of gravity
for a volume $\mathcal{V}$
reside in its boundary $\partial\mathcal{V}$} --- a  point of view that is strongly supported by the study
of horizon entropy, which shows that the degrees of freedom hidden by a horizon scales as the area and not as the volume. 
The resulting equations can be cast
in a thermodynamic form $TdS=dE+PdV$ and the
 continuum spacetime is like an elastic solid (see e.g. \cite{sakharov}]) with Einstein's equations providing the macroscopic description. Interestingly, the \cc\ arises again in this approach as a undetermined integration constant but closely related to the `bulk expansion' of the solid.

While this is all very interesting, we still need an extra physical principle to fix the value (even the sign) of \cc\ .
One possible way of doing this is to  interpret the $\Lambda$ term in the action as a Lagrange multiplier for the proper volume of the spacetime. Then it is reasonable to choose the \cc\ such that the total proper volume of the universe is equal to a specified number. While this will lead to a \cc\ which has the correct order of magnitude, it has several obvious problems. First, the proper four volume of the universe is infinite unless we make the spatial sections compact and restrict the range of time integration. Second, this will lead to a dark energy density  which varies as $t^{-2}$ (corresponding to $w= -1/3$ ) which is ruled out by observations. 

\subsection{Cosmic Lenz law}

Another possibility which has been attempted in the literature tries to ''cancel out'' the \cc\ by some process,
usually quantum mechanical in origin. One of the simplest ideas will be to ask whether switching on a \cc\ will
lead to a vacuum polarization with an effective energy momentum tensor that will tend to cancel out the \cc\ .
A less subtle way of doing this is to invoke another scalar field (here we go again!) such that it can couple to 
\cc\ and reduce its effective value \cite{lenz}. Unfortunately, none of this could be made to work properly. By and large, these approaches lead to an energy density which is either $\rho_{_{\rm UV}}\propto L_P^{-4}$ (where 
 $L_P$ is the Planck length) or to $\rho_{_{\rm IR}}\propto L_\Lambda^{-4}$ (where 
 $L_\Lambda=H_\Lambda^{-1}$ is the Hubble radius associated with the \cc\ ). The first one is too large while the second one is too small! 
 
 \subsection{Geometrical Duality in our Universe}
 
 While the above ideas do not work, it gives us a clue. A universe with two
 length scales $L_\Lambda$ and $L_P$  will be  asymptotically deSitter with $a(t)\propto \exp (t/L_\Lambda) $ at late times. There are some curious features in such a universe which we will now explore.  Given the two length scales $L_P$ and $L_\Lambda$, one can construct two energy scales
 $\rho_{_{\rm UV}}=1/L_P^4$ and $\rho_{_{\rm IR}}=1/L_\Lambda^4$ in natural units ($c=\hbar=1$). There is sufficient amount of justification from different theoretical perspectives
 to treat $L_P$ as the zero point length of spacetime \cite{zeropoint}, giving a natural interpretation to $\rho_{_{\rm UV}}$. The second one, $\rho_{_{\rm IR}}$ also has a natural interpretation. The universe which is asymptotically deSitter has a horizon and associated thermodynamics \cite{ghds} with a  temperature
 $T=H_\Lambda/2\pi$ and the corresponding thermal energy density $\rho_{thermal}\propto T^4\propto 1/L_\Lambda^4=
 \rho_{_{\rm IR}}$. Thus $L_P$ determines the \textit{highest} possible energy density in the universe while $L_\Lambda$
 determines the {\it lowest} possible energy density in this universe. As the energy density of normal matter drops below this value, the thermal ambience of the deSitter phase will remain constant and provide the irreducible `vacuum noise'. \textit{Note that the dark energy density is the the geometric mean $\rho_{_{\rm DE}}=\sqrt{\rho_{_{\rm IR}}\rho_{_{\rm UV}}}$ between the two energy densities.} If we define a dark energy length scale $L_{DE}$  such that $\rho_{_{\rm DE}}=1/L_{DE}^4$ then $L_{DE}=\sqrt{L_PL_\Lambda}$ is the geometric mean of the two length scales in the universe. (Incidentally, $L_{DE}\approx 0.04$ mm is macroscopic; it is also pretty close to the length scale associated with a neutrino mass of $10^{-2}$ eV; another intriguing coincidence ?!)
 
  \begin{figure}[ht]
  \begin{center}
  \includegraphics[angle=-90,scale=0.55]{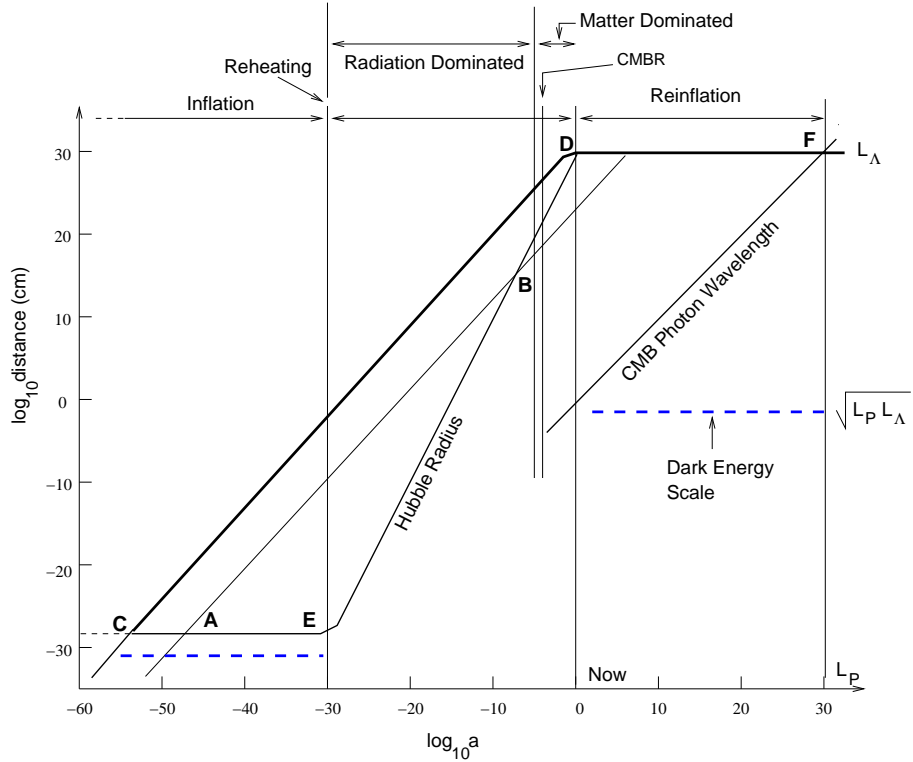}
  \end{center}
\caption{The geometrical structure of a universe with two length scales $L_P$ and $L_\Lambda$ corresponding to the Planck length and the cosmological constant \cite{plumian,bjorken}. Such a universe spends most of its time in two De Sitter phases which are (approximately) time translation invariant. The first De Sitter phase corresponds to the inflation and the second corresponds to the accelerated expansion arising from the cosmological constant. Most of the perturbations generated during the inflation will leave the Hubble radius (at some A, say) and re-enter (at B). However, perturbations which exit the Hubble radius
earlier than C will never re-enter the Hubble radius, thereby introducing a specific  dynamic range CE during the inflationary phase. The epoch F is characterized by the redshifted CMB temperature becoming equal to the De Sitter temperature $(H_\Lambda / 2\pi)$ which introduces another dynamic range DF in the accelerated expansion after which the universe is dominated by vacuum noise
of the De Sitter spacetime.}
\label{fig:tpplumian}
  \end{figure}
 
 Figure \ref{fig:tpplumian} summarizes these features \cite{plumian,bjorken}. Using the characteristic length scale of expansion,
 the Hubble radius $d_H\equiv (\dot a/a)^{-1}$, we can distinguish between three different phases of such a universe. The first phase is when the universe went through a inflationary expansion with $d_H=$ constant; the second phase is the radiation/matter dominated phase in which most of the standard cosmology operates and $d_H$ increases monotonically; the third phase is that of re-inflation (or accelerated expansion) governed by the cosmological constant in which $d_H$ is again a constant. The first and last phases are time translation invariant;
 that is, $t\to t+$ constant is an (approximate) invariance for the universe in these two phases. The universe satisfies the perfect cosmological principle and is in steady state during these phases!
 
 In fact, one can easily imagine a scenario in which the two deSitter phases (first and last) are of arbitrarily long duration \cite{plumian}. If  $\Omega_\Lambda\approx 0.7, \Omega_{DM}\approx 0.3$ the final deSitter phase \textit{does} last forever; as regards the inflationary phase, nothing prevents it from lasting for arbitrarily long duration. Viewed from this perspective, the in between phase --- in which most of the `interesting' cosmological phenomena occur ---  is  of negligible measure in the span of time. It merely connects two steady state phases of the universe.
 The figure \ref{fig:tpplumian} also shows the variation of $L_{DE}$ by broken horizontal lines. 
 
 While the two deSitter phases can last forever in principle, there is a natural cut off length scale in both of them
 which makes the region of physical relevance to be finite \cite{plumian}. Let us first discuss the case of re-inflation in the late universe. 
 As the universe grows exponentially in the phase 3, the wavelength of CMBR photons are being redshifted rapidly. When the temperature of the CMBR radiation drops below the deSitter temperature (which happens when the wavelength of the typical CMBR photon is stretched to the $L_\Lambda$.)
 the universe will be essentially dominated by the vacuum thermal noise of the deSitter phase.
 This happens at the point marked F when the expansion factor is $a=a_F$ determined by the
  equation $T_0 (a_0/a_{F}) = (1/2\pi L_\Lambda)$. Let $a=a_\Lambda$ be the epoch at which
  cosmological constant started dominating over matter, so that $(a_\Lambda/a_0)^3=
  (\Omega_{DM}/\Omega_\Lambda)$. Then we find that the dynamic range of 
 DF is 
 \begin{equation}
\frac{a_F}{a_\Lambda} = 2\pi T_0 L_\Lambda \left( \frac{\Omega_\Lambda}{\Omega_{DM}}\right)^{1/3}
\approx 3\times 10^{30}
\end{equation}

 Interestingly enough, one can also impose a similar bound on the physically relevant duration of inflation. 
 We know that the quantum fluctuations generated during this inflationary phase could act as seeds of structure formation in the universe \cite{genofpert}. Consider a perturbation at some given wavelength scale which is stretched with the expansion of the universe as $\lambda\propto a(t)$.
 (See the line marked AB in Figure \ref{fig:tpplumian}.)
 During the inflationary phase, the Hubble radius remains constant while the wavelength increases, so that the perturbation will `exit' the Hubble radius at some time (the point A in Figure \ref{fig:tpplumian}). In the radiation dominated phase, the Hubble radius $d_H\propto t\propto a^2$ grows faster than the wavelength $ \lambda\propto a(t)$. Hence, normally, the perturbation will `re-enter' the Hubble radius at some time (the point B in Figure \ref{fig:tpplumian}).
 If there was no re-inflation, this will make {\it all} wavelengths re-enter the Hubble radius sooner or later.
 But if the universe undergoes re-inflation, then the Hubble radius `flattens out' at late times and some of the perturbations will {\it never} reenter the Hubble radius ! The limiting perturbation which just `grazes' the Hubble radius as the universe enters the re-inflationary phase is shown by the line marked CD in Figure \ref{fig:tpplumian}. If we use the criterion that we need the perturbation to reenter the Hubble radius, we get a natural bound on the duration of inflation which is of direct astrophysical relevance. This portion of the inflationary regime is marked by CE
 and can be calculated as follows: Consider  a perturbation which leaves the Hubble radius ($H_{in}^{-1}$) during the inflationary epoch at $a= a_i$. It will grow to the size $H_{in}^{-1}(a/a_i)$ at a later epoch. 
 We want to determine $a_i$ such that this length scale grows to 
   $L_\Lambda$ just when the dark energy starts dominating over matter; that is at
 the epoch $a=a_\Lambda = a_0(\Omega_{DM}/\Omega_{\Lambda})^{1/3}$. 
  This gives 
  $H_{in}^{-1}(a_\Lambda/a_i)=L_\Lambda$ so that $a_i=(H_{in}^{-1}/L_\Lambda)(\Omega_{DM}/\Omega_{\Lambda})^{1/3}a_0$. On the other hand, the inflation ends at 
  $a=a_{end}$ where $a_{end}/a_0 = T_0/T_{\rm reheat}$ where $T_{\rm reheat} $ is the temperature to which the universe has been reheated at the end of inflation. Using these two results we can determine the dynamic range of CE to be 
  \begin{equation}
\frac{a_{\rm end} }{a_i} = \left( \frac{T_0 L_\Lambda}{T_{\rm reheat} H_{in}^{-1}}\right)
\left( \frac{\Omega_\Lambda}{\Omega_{DM}}\right)^{1/3}=\frac{(a_F/a_\Lambda)}{2\pi T_{\rm reheat} H_{in}^{-1}} \cong 10^{25}
\end{equation} 
where we have used the fact that, for a GUTs scale inflation with $E_{GUT}=10^{14} GeV,T_{\rm reheat}=E_{GUT},\rho_{in}=E_{GUT}^4$
we have $2\pi H^{-1}_{in}T_{\rm reheat}=(3\pi/2)^{1/2}(E_P/E_{GUT})\approx 10^5$.
If we consider a quantum gravitational, Planck scale, inflation with $2\pi H_{in}^{-1} T_{\rm reheat} = \mathcal{O} (1)$, the phases CE and DF are approximately equal. The region in the quadrilateral CEDF is the most relevant part of standard cosmology, though the evolution of the universe can extend to arbitrarily large stretches in both directions in time. 

This figure is definitely telling us something regarding the duality between Planck scale and Hubble scale or between the infrared and ultraviolet limits of the theory.
The mystery is compounded by the fact the asymptotic de Sitter phase has an observer dependent horizon and
related thermal properties. Recently, it has been shown --- in a series of papers, see ref.\cite{tpholo} ---  that it is possible to obtain 
classical relativity from purely thermodynamic considerations.  It is difficult to imagine that these features are unconnected and accidental; at the same time, it is difficult to prove a definite connection between these ideas and the \cc\ . Clearly, more exploration of these ideas is required.

\subsection{Gravity as detector of the vacuum energy}

Finally, I will describe an idea which \textit{does} lead to the correct value of \cc.
The conventional discussion of the relation between cosmological constant and vacuum energy density is based on
evaluating the zero point energy of quantum fields with an ultraviolet cutoff and using the result as a 
source of gravity.
Any reasonable cutoff will lead to a vacuum energy density $\rho_{\rm vac}$ which is unacceptably high. 
This argument,
however, is too simplistic since the zero point energy --- obtained by summing over the
$(1/2)\hbar \omega_k$ --- has no observable consequence in any other phenomena and can be subtracted out by redefining the Hamiltonian. The observed non trivial features of the vacuum state of QED, for example, arise from the {\it fluctuations} (or modifications) of this vacuum energy rather than the vacuum energy itself. 
This was, in fact,  known fairly early in the history of cosmological constant problem and, in fact, is stressed by Zeldovich \cite{zeldo} who explicitly calculated one possible contribution to {\it fluctuations} after subtracting away the mean value.
This
suggests that we should consider   the fluctuations in the vacuum energy density in addressing the 
cosmological constant problem. 

If the vacuum probed by the gravity can readjust to take away the bulk energy density $\rho_{_{\rm UV}}\simeq L_P^{-4}$, quantum \textit{fluctuations} can generate
the observed value $\rho_{\rm DE}$. One of the simplest models \cite{tpcqglamda} which achieves this uses the fact that, in the semiclassical limit, the wave function describing the universe of proper four-volume ${\cal V}$ will vary as
$\Psi\propto \exp(-iA_0) \propto 
 \exp[ -i(\Lambda_{\rm eff}\mathcal V/ L_P^2)]$. If we treat 
  $(\Lambda/L_P^2,{\cal V})$ as conjugate variables then uncertainty principle suggests $\Delta\Lambda\approx L_P^2/\Delta{\cal V}$. If
the four volume is built out of Planck scale substructures, giving $ {\cal V}=NL_P^4$, then the Poisson fluctuations will lead to $\Delta{\cal V}\approx \sqrt{\cal V} L_P^2$ giving
    $ \Delta\Lambda=L_P^2/ \Delta{\mathcal V}\approx1/\sqrt{{\mathcal V}}\approx   H_0^2
 $. (This idea can be a more quantitative; see \cite{tpcqglamda}).

Similar viewpoint arises, more formally, when we study the question of \emph{detecting} the energy
density using gravitational field as a probe.
 Recall that an Unruh-DeWitt detector with a local coupling $L_I=M(\tau)\phi[x(\tau)]$ to the {\it field} $\phi$
actually responds to $\langle 0|\phi(x)\phi(y)|0\rangle$ rather than to the field itself \cite{probe}. Similarly, one can use the gravitational field as a natural ``detector" of energy momentum tensor $T_{ab}$ with the standard coupling $L=\kappa h_{ab}T^{ab}$. Such a model was analysed in detail in ref.~\cite{tptptmunu} and it was shown that the gravitational field responds to the two point function $\langle 0|T_{ab}(x)T_{cd}(y)|0\rangle $. In fact, it is essentially this fluctuations in the energy density which is computed in the inflationary models \cite{inflation} as the seed {\it source} for gravitational field, as stressed in
ref.~\cite{tplp}. All these suggest treating the energy fluctuations as the physical quantity ``detected" by gravity, when
one needs to incorporate quantum effects.  
If the \cc\ arises due to the energy density of the vacuum, then one needs to understand the structure of the quantum vacuum at cosmological scales. Quantum theory, especially the paradigm of renormalization group has taught us that the energy density --- and even the concept of the vacuum
state --- depends on the scale at which it is probed. The vacuum state which we use to study the
lattice vibrations in a solid, say, is not the same as vacuum state of the QED.

 In fact, it seems \textit{inevitable} that in a universe with two length scale $L_\Lambda,L_P$, the vacuum
 fluctuations will contribute an energy density of the correct order of magnitude $\rho_{_{\rm DE}}=\sqrt{\rho_{_{\rm IR}}\rho_{_{\rm UV}}}$. The hierarchy of energy scales in such a universe, as detected by 
 the gravitational field has \cite{plumian,tpvacfluc}
 the pattern
 \begin{equation}
\rho_{\rm vac}={\frac{1}{ L^4_P}}    
+{\frac{1}{L_P^4}\left(\frac{L_P}{L_\Lambda}\right)^2}  
+{\frac{1}{L_P^4}\left(\frac{L_P}{L_\Lambda}\right)^4}  
+  \cdots 
\end{equation}  
 The first term is the bulk energy density which needs to be renormalized away (by a process which we  do not understand at present); the third term is just the thermal energy density of the deSitter vacuum state; what is interesting is that quantum fluctuations in the matter fields \textit{inevitably generate} the second term.

The key new ingredient arises from the fact that the properties of the vacuum state  depends on the scale at which it is probed and it is not appropriate to ask questions without specifying this scale. 
 If the spacetime has a cosmological horizon which blocks information, the natural scale is provided by the size of the horizon,  $L_\Lambda$, and we should use observables defined within the accessible region. 
The operator $H(<L_\Lambda)$, corresponding to the total energy  inside
a region bounded by a cosmological horizon, will exhibit fluctuations  $\Delta E$ since vacuum state is not an eigenstate of 
{\it this} operator. The corresponding  fluctuations in the energy density, $\Delta\rho\propto (\Delta E)/L_\Lambda^3=f(L_P,L_\Lambda)$ will now depend on both the ultraviolet cutoff  $L_P$ as well as $L_\Lambda$.  
 To obtain
 $\Delta \rho_{\rm vac} \propto \Delta E/L_\Lambda^3$ which scales as $(L_P L_\Lambda)^{-2}$
 we need to have $(\Delta E)^2\propto L_P^{-4} L_\Lambda^2$; that is, the square of the energy fluctuations
 should scale as the surface area of the bounding surface which is provided by the  cosmic horizon.  
 Remarkably enough, a rigorous calculation \cite{tpvacfluc} of the dispersion in the energy shows that
 for $L_\Lambda \gg L_P$, the final result indeed has  the scaling 
 \begin{equation}
 (\Delta E )^2 = c_1 \frac{L_\Lambda^2}{L_P^4} 
 \label{deltae}
 \end{equation}
 where the constant $c_1$ depends on the manner in which ultra violet cutoff is imposed.
 Similar calculations have been done (with a completely different motivation, in the context of 
 entanglement entropy)
 by several people and it is known that the area scaling  found in Eq.~(\ref{deltae}), proportional to $
L_\Lambda^2$, is a generic feature \cite{area}.
For a simple exponential UV-cutoff, $c_1 = (1/30\pi^2)$ but  cannot be computed
 reliably without knowing the full theory.
  We thus find that the fluctuations in the energy density of the vacuum in a sphere of radius $L_\Lambda$ 
 is given by 
 \begin{equation}
 \Delta \rho_{\rm vac}  = \frac{\Delta E}{L_\Lambda^3} \propto L_P^{-2}L_\Lambda^{-2} \propto \frac{H_\Lambda^2}{G}
 \label{final}
 \end{equation}
 The numerical coefficient will depend on $c_1$ as well as the precise nature of infrared cutoff 
 radius (like whether it is $L_\Lambda$ or $L_\Lambda/2\pi$ etc.). It would be pretentious to cook up the factors
 to obtain the observed value for dark energy density. 
 But it is a fact of life that a fluctuation of magnitude $\Delta\rho_{vac}\simeq H_\Lambda^2/G$ will exist in the
energy density inside a sphere of radius $H_\Lambda^{-1}$ if Planck length is the UV cut off. {\it One cannot get away from it.}
On the other hand, observations suggest that there is a $\rho_{vac}$ of similar magnitude in the universe. It seems 
natural to identify the two, after subtracting out the mean value by hand. Our approach explains why there is a \textit{surviving} cosmological constant which satisfies 
$\rho_{_{\rm DE}}=\sqrt{\rho_{_{\rm IR}}\rho_{_{\rm UV}}}$
 which ---  in our opinion --- is {\it the} problem. 
(For  a completely different way of interpreting this result, based on some imaginative ideas suggested by Bjorken, see
\cite{bjorken}).

\section{Conclusion}

In this talk I have argued that: (a) The existence of a component with negative pressure constitutes a major challenge in theoretical physics.
(b) The simplest choice for this component is the cosmological constant; other models based on scalar fields [as well as those based on branes etc. which I did not have time to discuss] do not alleviate the difficulties faced by \cc\  and --- in fact --- makes them worse. (c) The \cc\ is most likely to be a low energy relic of a quantum gravitational effect or principle and its explanation will require a radical shift in our current paradigm.
I discussed some speculative ideas and possible approaches to understand the \cc\ but none of them seems to be `crazy enough to be true'. Preposterous universe will require preposterous explanations and one needs to get bolder.

\end{document}